\documentclass[11pt,leqno]{amsart}
\usepackage{epsf,amsmath,amsfonts,amsthm,amssymb,epsfig,enumerate,setspace}
\newtheorem{thm}{Theorem}
\newtheorem{lem}[thm]{Lemma}
\newtheorem{prop}[thm]{Proposition}
\newtheorem{cor}[thm]{Corollary}
\newtheorem*{df}{Definition}

\newcommand{\qaa}{q_{\alpha,\alpha'}}
\newcommand{\E}{\mathbb{E}}

\newcommand{\prob}{\mathbb{P}}
\newcommand{\id}{\chi}
\newcommand{\sa}{\xi_\alpha}
\newcommand{\saprime}{\xi_{\alpha'}}

\newcommand{\qbar}{\bar{q}}

\newcommand{\ma}{\mathcal{M}_a}
\newcommand{\mak}{\mathcal{M}_{a,k}^{<1}}
\newcommand{\maone}{\mathcal{M}_{a}^{<1}}
\newcommand{\mc}{\mathcal{M}}
\newcommand{\Ppar}{\mathcal{P}_{\beta,h}}
\newcommand{\Ppargen}{\mathcal{P}_{\psi,g}}

\newcommand{\R}{\mathbb{R}}

\newcommand{\N}{\mathbb{N}}

\newcommand{\etat}{\tilde{\eta}}
\newcommand{\kappat}{\tilde{\kappa}}
\newcommand{\xit}{\tilde{\xi}}
\newcommand{\sat}{\tilde{\xi}_\alpha}

\newcommand{\F}{\mathcal{F}}

\newcommand{\Gpar}{G_{\beta,h}}
\newcommand{\eval}{\Big|}

\newcommand{\distrib}{\overset{\mbox{$\mathcal{D}$}}{=}}
\newcommand{\C}{\mathcal{C}}

\title{Spin Glass Computations and Ruelle's Probability Cascades}
\author{Louis-Pierre Arguin}\thanks{Work supported in part by NSERC and FQRNT postgraduate fellowships and NSF grant DMS 0602360}
\address{Department of Mathematics, Princeton University, Princeton NJ 08544}
\date{June 12th, 2006}

\begin{document}
\maketitle
%\tableofcontents
\begin{abstract}
We study the Parisi functional, appearing in the Parisi formula for the pressure of the SK model, as a functional on Ruelle's Probability Cascades (RPC). Computation techniques for the RPC formulation of the functional are developed. They are used to derive continuity and monotonicity properties of the functional retrieving a theorem of Guerra. We also detail the connection between the Aizenman-Sims-Starr variational principle and the Parisi formula. As a final application of the techniques, we rederive the Almeida-Thouless line in the spirit of Toninelli but relying on the RPC structure.
\end{abstract}

\section{Introduction}
The Sherrington-Kirkpatrick (SK) model is a mean-field spin glass system on configurations $\sigma\in\{\pm 1\}^N$ of $N$ spins with the Hamiltonian
$$ H_{N}(\sigma)=\frac{-1}{\sqrt{N}}\sum_{1\leq i<j\leq N}J_{ij}\sigma_i\sigma_j+h\sum_{1\leq i\leq N}\sigma_i.$$
The couplings $J_{ij}$ are independent standard gaussian variables and $h\in\R$.

It is now a theorem that the quenched pressure of the SK model in the thermodynamic limit, $P_{SK}(\beta,h):=\lim_{N\to\infty} \frac{1}{N}E_J\left[\log\sum_\sigma e^{-\beta H_{N}(\sigma)}\right]$, is given by the celebrated Parisi formula \cite{zeproof}:
$$ P_{SK}(\beta,h)=\inf_{x(\cdot)} \left\{\log 2 + f_x(0,h) -\frac{\beta^2}{2}\int_0^1qx(q)dq\right\}.$$
The infimum is over all increasing, right-continuous functions $x:q\mapsto x(q)$ on $[0,1]$ such that $x(0)=0$ and $x(1)=1$. At the heart of this formula is the so-called {\it Parisi functional} $x\mapsto f_x(0,h)$ where $f_x(q,y)$ is the solution to the partial differential equation
\begin{equation}
\partial_q f(q,y)+\frac{1}{2}\left[\partial^2_yf(q,y)+x(q) \left(\partial_y f(q,y)\right)^2\right]=0
\label{intro diff eqn}
\end{equation}
with the boundary condition $f(1,y)=\log\cosh(\beta y)$ \cite{Panchenko, Talagrand_parisi}. 

A different approach in computing the pressure of the SK model was taken by Aizenman, Sims and Starr ($AS^2$) \cite{ASS}. In this approach, the pressure is expressed through a general variational principle over {\it random overlap structures} (ROSt).  A ROSt is a measure $\mu$ on a pair $(\xi,Q)$ where $\xi=\{\xi_\alpha\}_{\alpha\in\mathcal{A}}$ is a set of weights labeled by $\mathcal{A}$ and $Q=\{\qaa\}$ is a positive semi-definite form on $\mathcal{A}$. The variational principle uses functionals on ROSt's of the form
\begin{equation}
E_\mu\left[\log\frac{\sum_{\alpha}\sa e^{\psi(\eta_\alpha)}}{\sum_{\alpha}\sa}\right]
\label{intro fct}
\end{equation}
where $\eta$ is a gaussian field with covariance $Q$ and $\psi$ is a specific function. 

As it was pointed out in \cite{ASS,ASS2}, the link between the two approaches in computing the SK pressure is provided by a particular family of ROSt's known as the {\it Ruelle's Probability Cascades} (RPC). In fact, the Parisi formula is retrieved by restricting the $AS^2$ variational problem to this class of ROSt's. These ROSt's possess two important features. First, their overlap matrix can be represented as a tree structure, sometimes qualified as {\it ultrametric}. Second, these processes are stable under stochastic shift of a certain kind. This property shall be defined precisely and be referred to as the {\it quasi-stationarity property}. 

The goal of this paper is to exhibit specific techniques for spin glass computations with the RPC's. The RPC is a natural setting to study the Parisi functional and its properties and to perform computations relevant to the SK model. 

The paper is organized as follows. We first introduce the RPC. We then precisely define the quasi-stationarity property and give sufficient and necessary conditions on the stochastic shift for stability. We proceed by studying functionals on RPC's of the form \eqref{intro fct} that reduce to Parisi-like functionals. We derive differentiation formulas for these functionals which naturally lead to continuity and monotonicity properties thereby retrieving a theorem of Guerra \cite{Guerra_bound} for the Parisi functional. Finally, we detail the connection between the $AS^2$ variational principle restricted to the class of RPC's and the Parisi formula for the pressure of the SK model. As an example of computation for this variational principle, we prove the instability of the high-temperature solution above the Almeida-Thouless line following the idea of Toninelli \cite{AT} but using the properties of the RPC functionals introduced earlier. It must be emphasized that the key element involved in most calculations is the quasi-stationarity property of the RPC.

 The connection between the RPC's and the SK model is still to be fully understood. In particular, the question of whether or not the overlap distribution of the SK model is supported on ultrametric matrices is open.
The quasi-stationarity property of the RPC seems to be at the core of this question as pointed out in \cite{ASS, ASS2,Guerra_cavity}. 

The author is thankful to Michael Aizenman for introducing him to the subject and for plenty of insightful discussions.

\section{ Ruelle's Probability Cascades}
\label{grem}
{\it Ruelle's Probability Cascades}, or RPC's, are cascades of Poisson point processes which carry a natural hierarchal distance between the atoms of the cascade. The RPC was formulated by Ruelle based on the {\it Generalized Random Energy Model} (GREM) originally defined by Derrida as a limit of finite point processes \cite{derrida}. Ruelle's formulation extends GREM's to allow continuous branchings or hierarchies \cite{Ruelle}. In these notes, we are interested in the case of finite number of branching levels where the definitions of GREM and RPC correspond. Keeping this in mind, we will often use the word GREM for an RPC with a finite number of branching levels.

%We will start by introducing the REM, the building block of the GREM. We will then define the class of GREM's, identifying each element of $\maone$ to a specific GREM process. %We will also look more closely at the distribution of the so-called overlaps between the points of the process following Bolthausen and Sznitman \cite{BS}. Finally, we will c%onstruct Gaussian fields on the cascade of the GREM which will enter in the definition of the GREM functionals of interest.

\subsection{Probability Measures on $[0,1]$}
We start by fixing the notation that will be needed in the definition of the GREM and used throughout the paper.

Let $\mc$ be the set of probability measures on $[0,1]$ and $\ma\subset\mc$, the subset of atomic measures with finite number of atoms.
 For later purposes, we also introduce $\mak$, the subspace of $\ma$ with exactly $k+1<\infty$ atoms, one of them located at $1$ and the remaining located on $[0,1)$. We write $\maone$ for $\bigcup_{k\in\N}\mak$. 
 
 The space $\mak$ corresponds to the following subspace of $[0,1]^{k+1}\times[0,1]^{k+1}$ most commonly used in the spin glass literature. We associate to $x\in\mak$ the pair $(\mathbf{x}, \mathbf{q})$ where $\mathbf{x}=(x_i,i=1,..,k+1)$ and $\mathbf{q}=(q_i,i=1,..,k+1)$ with the constraints
\begin{equation*}
\begin{aligned}
0&< x_1 < x_2 < ... < x_{k+1}\equiv 1 \\
0&\leq q_1 < q_2 < ... <q_{k+1}\equiv 1.
\end{aligned}
\end{equation*}
We also set $q_0=0$. In this notation, $q_i$ refers to the position of the $i$-th atom and $x_i=x(q_i)$.

Throughout these notes, we will identify a probability measure with its distribution function and write $x\in\mc$ for a distribution function $x$ of a measure in $\mc$. We will be naturally led to endow $\mc$ with the topology induced by the $L^1([0,1],g'(q)dq)$-norm on the distribution functions for a given smooth function $g$.  It turns out that the topology does not actually depend on $g$. We refer to this topology as the $L^1$-topology on $\mc$. In fact, the $L^1$-topology is simply the weak topology on $\mc$ (see Appendix \ref{app:topo} for details). Note that the subset $\maone$ is dense in $\mc$ in the $L^1$-topology.

\subsection{The REM}
The building block of the cascade of the GREM is the simple REM point process.
\begin{df}[REM]
Let $0<x<1$. A REM($x$) is a Poisson point process on $\R^+$ with intensity measure $xs^{x-1}ds$.
\end{df}
The value of the parameter $x=1$ is evidently singular. It will become clear later that this is the fundamental reason for introducing the subspace $\maone\subset \ma$. Fortunately, we will later consider functionals of the point process that allow a continuous extension to the case $x=1$. 

Let $\zeta=\{\zeta_\alpha\}$ be a REM($x$). It is not too hard to show that $\sum_\alpha\zeta_\alpha<\infty \text{ a.s.}$ (see e.g. \cite{Ruelle}). In particular, $\zeta$ is locally finite on $(0,\infty)$ and bounded on the right almost surely. Therefore, it is possible to enumerate the points of a realization in decreasing order i.e. $\zeta_1>\zeta_2>...$. 

It turns out that the REM possesses an interesting stability property under stochastic shift. This property is at the root of the techniques and results presented in this paper. Let $\zeta$ be a REM($x$). We consider a random variable $W$ on $\R^+$ with distribution $\nu$ such that $E_\nu[W^x]=\int_0^\infty w^xd\nu(w) <\infty.$ A proof of the following can be found in Proposition 3.1 of \cite{RA}.
\begin{prop}[Quasi-Stationarity of the REM]
Let $\zeta=\{\zeta_i\}_{i\in\N}$ be a REM($x$) and $W$ be as above. Consider $\{W_i\}_{i\in\N}$ iid $W$-distributed and independent of $\zeta$. Define the point process $ \tilde{\zeta}:=\{\zeta_i W_i\}_{i\in\N}.$
The following hold
\begin{enumerate}
\item \underline{Quasi-Stationarity}: $\tilde{\zeta}$ is a REM($x$) scaled by $E_\nu[W^x]^{1/x}$, i.e. $$\tilde{\zeta}\distrib E_\nu[W^x]^{1/x}\zeta.$$
\item  \underline{Backward Shift}: Let $\{\tilde{\zeta_j}\}$ be ordered in decreasing order. Let $\pi:\N\to\N$ be the random permutation induced by the random shift, i.e. $\pi(i)=j$ iif $\tilde{\zeta}_j=\zeta_i W_i$. Then, $\{W_{\pi^{-1}(j)}\}_{j\in\N}$ are iid and independent of $\tilde{\zeta}$ with distribution 
$$ \frac{w^xd\nu(w)}{E_\nu[W^x]^{1/x}}.$$
\end{enumerate}
\label{REM QS}
\end{prop}

\subsection{The GREM}
We now construct the GREM process as a cascade or hierarchy of REM's. We start by defining the point process associated to a GREM, we then introduce the overlap matrix induced by the cascade. We choose to identify the class of GREM processes with the space of atomic measures $\maone$. Therefore, elements of the class of GREM's are distinguished by the choice of $x_l$'s but also by the choice of overlap parameters $q_l$'s.  This is a useful identification as functionals over GREM's become functionals on a dense subspace of the space $\mc$ of probability measures on $[0,1]$. 

Let $x\in\mak$ with atoms at $\{q_l\}_{1\leq l\leq k+1}$ and $x(q_l)=x_l$. Recall that $q_{k+1}=1$ and $x_{k+1}=1$. 
Consider $\alpha\in\N^{k}$, $\alpha=(\alpha_1,...,\alpha_{k})$.
It is convenient to define, for $l=0,...,k$, the truncation $\alpha(l):=(\alpha_1,...,\alpha_l)$. By convention, $\alpha(0)=0$. 
Consider for each $l=1,...,k$ a collection of independent REM($x_{l}$) indexed by $\alpha(l-1)\in\N^{l-1}$
\begin{equation*}
\left(\zeta^{\alpha(l-1)},\alpha(l-1)\in\N^{l-1}\right).
\label{zeta}
\end{equation*}
The notation $\left(\zeta^{\alpha(l-1)}\right)_j$ will designate the $j$-th point of a realization of the process $\zeta^{\alpha(l-1)}$.
By the convention $\alpha(0)=0$, there is only one process in the collection $l=1$. 
%For the singular case $l=k$ for which $x_k=1$, we pick $\zeta_{\alpha(l)}$ as a sequence of iid random variables on $\R^+$. 

We define recursively a hierarchy of point processes $\{\xi_{\alpha(l)}\}_{\alpha(l)\in\N^l}$ at levels $0\leq l\leq k$ as follows: $\xi_{\alpha(0)}:= 1$ and $\xi_{\alpha(l)}:= \xi_{\alpha(l-1)}\left(\zeta^{\alpha(l-1)}\right)_{\alpha_l}$. The resulting point process is then
\begin{equation*}
\xi_{\alpha}:= \xi_{\alpha(k-1)}\left(\zeta^{\alpha(k-1)}\right)_{\alpha_k}.
\label{xi_trunc}
\end{equation*}
%We write $\Omega^\xi$ be the probability space on which all the processes $\xi_{\alpha(l)}$ are defined. 
To keep track on the branching information, it is useful to consider the filtration $\F^\xi=(\F^\xi_l,0\leq l\leq k)$ with $\F_0$ being the trivial $\sigma$-algebra and $\F_l^\xi=\sigma\left(\{\xi_{\alpha(l')}\}_{\alpha(l')\in\N^{l'}},0\leq l'\leq l\right)$. Here, $\sigma(\cdot)$ designates the $\sigma$-algebra generated by the collection of variables therein.

The overlap matrix of the cascade $Q=\{\qaa\}_{\alpha,\alpha'\in\N^k}$ is defined as
\begin{equation*}
q_{\alpha,\alpha'}:= \max\{q_{l+1}: \text{for $l$ such that $\alpha(l)=\alpha'(l)$}\}.
\end{equation*}
The overlap matrix is clearly symmetric. Also, if $\alpha(1)\neq \alpha'(1)$, then $q_{\alpha,\alpha'}=q_1$ and $q_{\alpha,\alpha}=q_{k+1}=1$.
In addition, the following inequality holds by definition
\begin{equation}
q_{\alpha,\alpha'}\geq \min\{q_{\alpha,\alpha''};q_{\alpha',\alpha''}\}
\label{um3}
\end{equation}
for any triplet $\alpha,\alpha',\alpha''$.
For $d_{\alpha,\alpha'}:=1-q_{\alpha,\alpha'}$, the inequality becomes the {\it ultrametric inequality}: $ d_{\alpha,\alpha'}\leq\max\{ d_{\alpha,\alpha''},d_{\alpha',\alpha''}\}$. It implies that at least two overlaps in the triplet must be the same, and the distinct one, if any, must be greater than the redundant overlap.
As we will see later in the construction of the cavity field, the overlap matrix is also the covariance matrix of a gaussian field labeled by $\alpha\in\N^k$. In particular, it is positive definite.

\begin{df}[GREM]
Let $x\in\maone$ with atoms at $\{q_l\}_{1\leq l\leq k}$ and $q_{k+1}=1$ with $x(q_l)=x_l$ and $x(q_{k+1})=1$. A GREM($x$) is the pair $(\xi,Q)$ where $\xi=\{\xi_\alpha\}_{\alpha\in\N^k}$ is the resulting point process 
%on $\Omega^\xi$
 constructed above with parameters $x_l$, $1\leq l\leq k$, and $Q=\{q_{\alpha,\alpha'}\}$ is the corresponding symmetric, positive definite matrix with $q_{\alpha,\alpha'}\in\{q_1,...,q_k,1\}$. Note that, by definition, a GREM($x$) is a ROSt.
\end{df}
We stress that the letter $\xi$ will be used for the resulting process $\xi=\{\xi_\alpha\}_{\alpha\in\N^k}$ and not for the whole cascade of processes. In particular, $\xi$ does not contain information on the hierarchy. The information on the hierarchy is encoded in the labeling $\alpha$ and expressed through the overlap matrix $Q$.

The point process $\xi=\{\xi_\alpha\}_{\alpha\in\N^k}$ keeps some regularity features of the REM. Indeed, $\sum_\alpha \sa<\infty$ $a.s.$ and $\xi$ is a random Poisson process whose intensity measure, conditioned on $\F^\xi_{k-1}$, is $\sum_{\alpha(k-1)\in\N^{k-1}}\xi^{x_k}_{\alpha(k-1)}x_k s^{x_k-1}ds.$ These facts are consequences of basic properties of the REM. Proofs can be found in \cite{BS}, Lemma 2.1.

As for the REM, the summability allows the enumeration of the points of a realization in decreasing order, i.e. $\{\xi_\alpha\}_{\alpha\in\N^k}=\{\xi_i\}_{i\in\N}$, $\xi_1>\xi_2>...$. This ordering induces a random bijection $\phi: \N\to \N^{k} $ where $\phi(i)=\alpha$ if $\xi_\alpha=\xi_i$. The matrix $\phi^{-1}\circ Q\circ \phi=\{q_{\phi(i)\phi(j)}\}_{i,j\in\N}$ is now clearly random. We will sometimes abuse notation and write $q_{ij}$ for $q_{\phi(i)\phi(j)}$ and $Q$ for $\phi^{-1}\circ Q\circ \phi$. The intended meaning will be clear from the notation and the context. The distribution of the matrix $\phi^{-1}\circ Q\circ \phi$ looks intricate at first due to its dependence on the ordering of the process $\xi$. It turns out it has a simple form due to Bolthausen and Sznitman \cite{BS}.

To illustrate this distribution, we must define the following random equivalence relations on $\N$ for each level $l$, $0\leq l\leq k$:  
\begin{equation}
i\sim_l j \text{ if and only if $[\phi(i)](l)=[\phi(j)](l)$.}
\label{relations}
\end{equation}
By convention, $i\sim_0 j$ for all $i,j\in\N$. We write $\Gamma_{x_l}$ for the partition of $\N$ induced by $\sim_l$, i.e. $\Gamma_{x_l}:=\N/\sim_l$. $\Gamma_{x_l}$ is obtained by lumping equivalence classes of $\Gamma_{x_{l+1}}$. This is because $i\sim_{l} j$ if  $i\sim_{l+1} j$ by the definition \eqref{relations}. The distribution of the sequence of partitions $\{\Gamma_{x_l}\}$ is surprisingly simple.

\begin{thm}[Theorems 1.2, 2.2 and Proposition 1.4 in \cite{BS}]
Define $\Gamma(t):=\Gamma_{e^{-t}}$.
The process $\Gamma(t)$, $t=0,-\log x_k, ...,-\log x_1<\infty$, is a discrete-time Markov process on the space of partitions of $\N$ whose transition probabilities are defined as follows.

Consider $\Gamma^{(n)}(s)$ and $\Gamma^{(n)}(t)$, sets of equivalence classes of $\{1,....,n\}$. Define $k_s=|\Gamma^{(n)}(s)|$ and $k_t=|\Gamma^{(n)}(t)|$. Let  $\Gamma^{(n)}(t)$ be obtained from $\Gamma^{(n)}(s)$ by respectively lumping $m_1$,..., $m_{k_t}$ classes of $\Gamma^{(n)}(s)$. Then the transition probability from $\Gamma^{(n)}(s)$ to $\Gamma^{(n)}(t)$, $s<t$, is
$$ P(s\to t ;\Gamma^{(n)}(s),\Gamma^{(n)}(t))= \frac{(k_t-1)!}{(k_s-1)!}\left(\frac{e^{-t}}{e^{-s}}\right)^{(k_t-1)}\prod_{l=1}^{k_t}u(m_l,e^{-t}/e^{-s})$$
where $u(1,x)=1$ and $u(m,x)=\frac{\left(m-1-x\right)...\left(1-x\right)}{(m-1)!}.$
\label{prob_grem}

Moreover, the process $\Gamma(t)$ is independent from the normalized point process $\{\xi_i/\sum_i\xi_i\}_{i\in\N}$.
\end{thm}

Given a realization of the equivalence relations $\sim_l$, the overlap of the $i$-th and $j$-th points follows from equation \eqref{relations}:
$$ q_{ij}:=q_{\phi(i)\phi(j)}=q_{\max\{l+1: \text{ $i\sim_l j$}\}}$$ 
Clearly, the distribution of the process $Q=\{q_{ij}\}$ on symmetric, positive definite matrices depends uniquely on the distribution of the random equivalence classes given above. As an example of calculation of overlap probabilities using Theorem \ref{prob_grem}, we have $\prob_x(q_{12}=q_l)=(x_{l}-x_{l-1})$. It is important to note that the last assertion of Theorem \ref{prob_grem} implies that the process $Q$ is independent from the normalized weights $\{\xi_i/\sum_i\xi_i\}_{i\in\N}$.
 
\subsection{The Cavity Field of the GREM}   
In this section, we introduce the gaussian fields on the GREM that will appear in the definition of the functionals of interest on RPC's.
\begin{df}[Gaussian Field on the GREM]
Let $(\xi,Q)$ be a GREM($x$) and $g:[0,1]\to \R^+$, a strictly increasing function in $C^1([0,1])$ with $g(0)=0$.
A {\it gaussian field with covariance function $g$ on $\xi$} is a centered gaussian process $\kappa=(\kappa_\alpha(r),\alpha\in\N^k,0\leq r\leq 1)$ with covariance
$$\text{Cov}(\kappa_\alpha(q),\kappa_{\alpha'}(q'))=\int_0^{q\wedge q'\wedge q_{\alpha,\alpha'}} g'(r) dr= g(q\wedge q'\wedge q_{\alpha,\alpha'}).$$
In the case $g(q)=q$, the field is called the cavity field of $\xi$ and will be denoted by $\eta$.  
We will write $\kappa_i$ for $\kappa_{\phi(i)}$, the field of the $i$-th point of the configuration.
%We denote by $\Omega^\kappa$ the probability space on which the cavity field is defined.
\end{df}

The cavity field $\eta$ can be constructed explicitly. We consider, for each $\alpha(l)\in\N^l$ and each $0\leq l\leq k$, independent standard brownian motions $B_{\alpha(l)}$ on $[q_l,q_{l+1})$. One could think of the brownian motion as attached to each subbranch $ \alpha(l)$ of the cascade. For each $\alpha\in\N^k$, we construct the process $\eta_\alpha$ recursively. We set $\eta_{\alpha(0)}(q)=B_{\alpha(0)}$ for $0\leq q \leq q_1$ and $\eta_{\alpha(l)}(q)=\eta_{\alpha(l-1)}(q_l) + B_{\alpha(l)}(q)$ for $q_l\leq q \leq q_{l+1}$.
Finally, $\eta_{\alpha}(q)=\eta_{\alpha(l)}(q)$ where $q_l\leq q \leq q_{l+1}$.
It is straightforward to check that $\text{Cov}(\eta_\alpha(q),\eta_{\alpha'}(q'))=q\wedge q'\wedge q_{\alpha,\alpha'}$.
In particular, $\text{Cov}(\eta_\alpha(1),\eta_{\alpha'}(1))=q_{\alpha,\alpha'}$ which shows that the overlap matrix $Q$ is positive definite as claimed before. The gaussian field $\kappa$ with covariance function $g$ can now be written as a stochastic integral on $\eta$
$$\kappa_\alpha(q)=\int_0^q \sqrt{g'(r)} d\eta_{\alpha}(r).$$

The natural filtration of the cavity field on the tree is $\F^\eta_q=\sigma\left(\eta_{\alpha}(r),\alpha\in\N^k ,0\leq r\leq q \right).$
It is useful to construct the filtration of a GREM($x$) which keeps track of the information of the cascade of point processes as well as the cavity field on it. Let $\F=(\F_q,q\in [0,1])$ be the right-continuous filtration defined by $\F_q=\F_{l-1}^\xi \bigotimes \F_q^\eta $ for $q\in [q_l,q_{l+1})$ and for $q=1$, $\F_1=\F_{k}^\xi \bigotimes \F_1^\eta.$
Note that the filtration is actually continuous at every point except at the points $q_l$ where branchings occur. Also, the $\sigma$-algebra $\F_q$ contains the information of the branchings strictly above $[q_l,q_{l+1})$ for $q$ in this interval as symbolized by $\F^\xi_{l-1}$.
\begin{df}
Throughout the rest of this paper, we will write 
%$\Omega$ for $\Omega^\xi\times\Omega^\eta$ and 
$\prob_x$ for the probability measure on $\F_1$ of a GREM($x$) and its cavity field. $\E_x$ will denote the expectation. 
\end{df}

%At this point, it is useful to sum up the GREM construction in Figure \ref{fig:grem_tree}. 

%\begin{figure}[htbp]
%\begin{center}
%\leavevmode
%\includegraphics[clip,width = 5cm]{grem_tree.eps}
%\end{center}
%\caption{Illustration of a GREM tree and its cavity field (built upon the brownian motions $B_{\alpha(l)}$) in the case $k=2$. Each Poisson process will produce an infinite number of points in reality. The bijection $\phi$ of the realization yields for example $\phi(4)=(3,1)$.}
%\label{fig:grem_tree}
%\end{figure}

\section{The Quasi-Stationarity Property of the GREM}
The quasi-stationarity property of the REM stated in Proposition \ref{REM QS} induces a similar stability property on a cascade of REM. We now study this important feature of the GREM process. The stochastic shift at each point will be written as a function of the gaussian field on the RPC presented in the last section. We also look at the distribution of the field after a shift. It is modified by the reordering as it was in the REM case.

We first need to introduce a class of function for which the stochastic shift is well-defined.
\begin{df}
Let $\mathcal{C}$ be the class of functions $\psi$ in $C^2(\R)$ satisfying
\begin{itemize}
\item $\psi'$ and $\psi''$ are bounded on $\R$;
\item $E_Y\left[e^{\psi(Y)}\right]<\infty$ for any gaussian variable $Y$.
\end{itemize} 
\end{df}
We remark that $E_Y\left[e^{x\psi(Y)}\right]< \infty$ for any $0<x<1$ (using Jensen's inequality applied with the convex function $f(y)=y^{1/x}$). Also, $\mathcal{C}$ includes the functions $\psi(\kappa)=\log\cosh(\beta\kappa+h)$ and $\psi(\kappa)=\beta\kappa$. 

In this section, we are interested in the stability properties of the GREM under the stochastic shift
\begin{equation*}
\sa\mapsto \sa e^{f_q(\kappa_\alpha(q))}
\end{equation*}
where $\kappa$ is a gaussian field on $\xi$ with covariance function $g$. The family of functions $(f_q,q\in[0,1])$ is assumed to be contained in the class $\C$. We also assume that the family is differentiable, i.e. for $y$ fixed, $f_q(y)$ is a differentiable function of $q$.  

An example of such a family is the following. Fix $x\in\maone$. Let $\psi\in \C$ and $\psi_1:=\psi$. For $q\in[q_l,q_{l+1})$, we define recursively $\psi_q$ as
\begin{align}
\psi_q(y)&=\frac{1}{x_l}\log E_z\left[e^{x_l\psi_{q_{l+1}}(y+z\sqrt{g(q_{l+1})-g(q)})}\right]\text{  for $1\leq l\leq k$};\label{psiq}\\
\psi_{q}(y)&=\log E_z\left[e^{\psi_{q_{l+1}}(y+z\sqrt{g(q_{1})-g(q)})}\right] \text{  for $l=0$}\nonumber
\end{align}
where $E_z$ denotes the expectation over $z$, a standard gaussian. We dropped the dependence of $\psi_q$ on $x$ and $g$ but the reader must keep in mind this dependence. It is easy to verify that $\psi_q$ is in $\C$. Moreover, by direct derivation, $\psi_q$ is seen to satisfy the differential equation (see e.g. \cite{zeproof})
\begin{equation*}
\partial_q\psi_q(y)+\frac{g'(q)}{2}\left(\psi''_q(y)+x_l(\psi'_q(y))^2\right)=0
\end{equation*}
with the continuity condition $\lim_{q\to q^-_{l+1}}\psi_q(y)=\psi_{q_{l+1}}(y)$.
More generally, by using the condition of continuity and the equation for each interval, the function $\psi_q(y)$ actually satisfies
\begin{equation*}
\partial_q\psi_q(y)+\frac{g'(q)}{2}\left(\psi''_q(y)+x(q)(\psi'_q(y))^2\right)=0
\label{diff eqn}
\end{equation*}
with boundary condition $\psi_1(y)=\psi(y)$.

As a matter of fact, the function $e^{\psi_q(y)}$ is the factor $E[W^{x_l}]^{1/x_l}$ coming from the application of the quasi-stationarity property of Proposition \ref{REM QS} to a REM($x_l$) with a shift $W_\alpha=e^{\psi_{r}(y+Y_\alpha)}$, $q<r\leq q_{l+1}$, where $Y_\alpha$ are independent gaussians $\mathcal{N}(0,g(r)-g(q))$.  

The family of functions $\psi_q$ has the property that the GREM distribution under their associated stochastic shift is invariant up to a random common factor (labeled by $\alpha(0)$):
\begin{equation}
(\xi_\alpha e^{\psi_q(\kappa_\alpha(q))},Q)\distrib \left(\xi_\alpha e^{\psi_{q_1}(\kappa_{\alpha(0)}(q_1))},Q\right)
\label{stab1}
\end{equation}
for $q\in[q_1,1]$. Indeed, let $q\in[q_l,q_{l+1}]$. We write $\delta_{\alpha(l)}(q):=\kappa_{\alpha(l)}(q)-\kappa_{\alpha(l)}(q_l)$. Note that the $\delta_{\alpha(l)}$'s are independent gaussian $\mathcal{N}(0,g(q)-g(q_l))$ for each $\alpha(l)$. Therefore, we can apply the quasi-stationarity to each REM $\zeta^{\alpha(l-1)}$:
\begin{align*}
\xi_\alpha e^{\psi_q(\kappa_\alpha(q))}&=\xi_{\alpha(l-1)}\left(\zeta^{\alpha(l-1)}\right)_{\alpha_l}e^{\psi_q(\kappa_\alpha(q_l)+\delta_{\alpha(l)}(q))}\prod_{l'=l+1}^k\left(\zeta^{\alpha(l'-1)}\right)_{\alpha_l'}\\
&\distrib \xi_{\alpha(l-1)}\left(\zeta^{\alpha(l-1)}\right)_{\alpha_l}E_{\delta_{\alpha(l)}}\left[e^{x_l\psi_q(\kappa_{\alpha}(q_l)+\delta_{\alpha(l)}(q))}\right]^{1/x_l}\prod_{l'=l+1}^k\left(\zeta^{\alpha(l'-1)}\right)_{\alpha_l'}\\
&=\xi_\alpha e^{\psi_{q_l}(\kappa_\alpha(q_l))}
\end{align*}
where we used the definition of $\psi_{q_l}$ in the last equality. This procedure is applied successively up to $q_1$ to prove the claim.

Surprisingly, it turns out that the property \eqref{stab1} characterizes the family $(\psi_q,q\in[0,1])$.
\begin{thm}[Quasi-Stationarity of the GREM]
Let $(f_q,q\in[0,1])$ be a differentiable family of functions in $\C$. Let $(\xi,Q)$ be a GREM($x$) and $\kappa$, its gaussian field with covariance function $g$. Then the function $f_q(y)$ satisfies the differential equation \eqref{diff eqn} for $q\in[q_1,1]$ if and only if
\begin{equation*}
(\xi_\alpha e^{f_q(\kappa_\alpha(q))},Q)\distrib \left(\xi_\alpha e^{f_{r}(\kappa_\alpha(r))},Q\right)
\label{stability}
\end{equation*}
for all $q,r\in[q_1,1]$. In particular,
\begin{equation}
(\xi_\alpha e^{f_1(\kappa_\alpha(1))},Q)\distrib \left(\xi_\alpha e^{f_{q_1}(\kappa_{\alpha(0)}(q_1))},Q\right).
\label{QS fct}
\end{equation}
\label{GREM QS}
\end{thm}
\begin{proof}
The sufficiency of the differential equation for stability was proven above. We prove the necessity. Pick $q\in[q_l,q_{l+1})$, $l\geq 1$. Choose $\Delta q$ small enough so that $q+\Delta q$ belongs also to $[q_l,q_{l+1})$. Consider $\delta_{\alpha(l)}(\Delta q):=\kappa_{\alpha(l)}(q+\Delta q)-\kappa_{\alpha(l)}(q)$ that are independent gaussian $\mathcal{N}(0,g(q+\Delta q)-g(q))$ for each $\alpha(l)$. Applying quasi-stationarity, we obtain 
$$\xi_\alpha e^{f_{q+\Delta q}(\kappa_\alpha(q)+\delta_{\alpha(l)}(\Delta q))}\distrib\xi_\alpha E_{\delta_{\alpha(l)}}\left[e^{x_lf_{q+\Delta q}(\kappa_\alpha(q)+\delta_{\alpha(l)}(\Delta q))}\right]^{1/x_l}.$$
But, by equation \eqref{stability}, we also have
$$\xi_\alpha e^{f_{q+\Delta q}(\kappa_\alpha(q)+\delta_{\alpha(l)}(\Delta q))}\distrib\xi_\alpha e^{f_{q}(\kappa_\alpha(q))}.$$
As this must hold for all realization of $\kappa_\alpha(q)$, we conclude that $E_{\delta_{\alpha(l)}}\left[e^{x_lf_{q+\Delta q}(y+\delta_{\alpha(l)}(\Delta q))}\right]=e^{x_lf_q(y)}$ for all $\Delta q$. In particular,
$$ \lim_{\Delta q\to 0}\frac{1}{\Delta q}\left(E_{\delta_{\alpha(l)}}\left[e^{x_lf_{q+\Delta q}(y+\delta_{\alpha(l)}(\Delta q))}\right]-e^{x_lf_{q}(y)}\right)=0$$
if the limit exists. The limit does exist and is easily computed by It\^o's formula
$$\frac{d}{dr}E_{\delta_{\alpha(l)}}\left[e^{x_lf_{r}(y+\delta_{\alpha(l)}(r))}\right]\eval_{r=0}=x_le^{x_lf_q(y)}\left(\partial_q f_q(y)+\frac{g'(q)}{2}(f''_q(y)+x_l(f'_q(y))^2)\right)=0$$
which yields the desired differential equation.
\end{proof}

From now on, we will use the notation $\xit_\alpha$ for the shifted process $\xi_\alpha e^{\psi(\kappa_{\alpha}(1))}$.
Consider the random permutation $\pi$ of $\N$ induced by this random shift, i.e. $\pi(i)=i'$ if $\xit_{i'}=\xi_i e^{\psi(\kappa_i(1))}$. As in Proposition \ref{REM QS}, we are interested in the backward distribution of the random shift. More precisely, we study the distribution of the field $\kappat$:
 $$\kappat:=(\kappat_i,i\in \N)=(\kappa_{\pi^{-1}(i)},i\in \N).$$

\begin{prop}[Backward Distribution of the field]
Let $\mu^{(n)}_Q$ be the joint distribution of the gaussian fields $\kappa_1$,...,$\kappa_n$ under $\prob_x$ given the overlap $Q$ and $\tilde{\mu}^{(n)}_Q$, the distribution of $\kappat_1$,...,$\kappat_n$ under $\prob_x$ given $Q$. Then $\tilde{\mu}^{(n)}_Q$ is absolutely continuous with respect to $\mu^{(n)}_Q$ and
$$\frac{d\tilde{\mu}^{(n)}_Q}{d\mu^{(n)}_Q}=\prod_{l=0}^{k}\prod_{\bar{i}\in\{1,...,n\}/\sim_l}V_{\bar{i}} $$
where the product is over the equivalence classes of $\{1,...,n\}/\sim_l$ and
\begin{equation*}
V_{\bar{i}}=\frac{e^{x_l\psi_{q_{l+1}}(\kappat_{i}(q_{l+1}))}}{e^{x_l\psi_{q_{l}}(\kappat_{i}(q_{l}))}}.
\label{V}
\end{equation*}
\label{backward cavity}
 \end{prop}
From the form of the distribution, we conclude that $\kappat$ is independent from $\xit$. The last proposition seems abstract at first but will prove extremely useful as it will allow us to express derivatives of the Parisi functional (and more generally, of functionals on RPC's) in a compact way as expectations over the backward gaussian field.
\begin{proof}
We prove the case $n=1$. Choose a finite set of $t_j$'s in $[0,1]$, $j=1,...,J$. We consider the union of $\{t_j\}_{j\in J}$ and $\{q_l\}_{1\leq l\leq k+1}$: $\{r_i\}_{i=1,...,N}=\{t_j\}\cup \{q_l\}$ so $r_1<r_2<...<r_N$.
The claim will be proven for $n=1$ if for any choice of $t_j$, the finite-dimensional distributions satisfy
$$ \tilde{\mu}^{(1)}(\kappat_1(t_j)\in A_j, 1\leq j \leq J)= \int_{B_1}...\int_{B_N} \prod_{l=0}^{k}
\frac{e^{x_l\psi_{q_{l+1}}(\kappat_{1}(q_{l+1}))}}{e^{x_l\psi_{q_{l}}(\kappat_{1}(q_l))}}\mu^{(1)}(\kappat_{1}(r_i)\in ds_i, 1\leq i\leq N).$$
where $B_i=A_j$ if $r_i=t_j$ and $B_i=\R$ otherwise. 

Define $l_i$ as the level $l$ where $r_i\in (q_l,q_{l+1}]$. By definition, if $r_i\neq q_l$ for any $l$ then $l_{i+1}=l_i$. The key point in the definition of the $r_i$ is that the $\kappa_{\alpha(l_i)}$'s, given $\F_{r_{i-1}}$, are independent. (This would not be true if one does not consider all $q_l$'s in the definition of $r_i$'s.)

By Theorem \ref{GREM QS}, we have that, 
$$\xit_{\alpha(l_i)}= \xi_{\alpha(l_i)}e^{\psi_{r_i}(\kappa_{\alpha(l_i)}(r_i))}\distrib\xi_{\alpha(l_i)}e^{\psi_{r_{i-1}}(\kappa_{\alpha(l_i)}(r_{i-1}))}.$$ 
By the quasi-stationarity property of the REM (Proposition \ref{REM QS}), the backward distribution of $\kappa_{1}(r_i)$ given $\F_{r_{i-1}}$ is
\begin{equation*}
\frac{e^{x_{l_i}\psi_{r_i}(\kappat_{1}(r_i))}}{e^{x_{l_i}\psi_{r_{i-1}}(\kappat_{1}(r_{i-1}))}}\mu^{(1)}(\kappat_{1}(r_i)\in ds | \F_{r_{i-1}}).
\label{V proof}
\end{equation*}
The joint distribution of the $\kappat_{1}(r_i)$ is therefore
$$ \tilde{\mu}^{(1)}(\kappat_1(r_i)\in B_i, 1\leq i\leq N)=\int_{B_1}...\int_{B_N}\prod_{i=1}^N\frac{e^{x_{l_i}\psi_{r_i}(\kappat_{1}(r_i))}}{e^{x_{l_i}\psi_{r_{i-1}}(\kappat_{1}(r_{i-1}))}}\mu^{(1)}(\kappat_{1}(r_i)\in ds_i, 1\leq i\leq N).$$
If $r_i\neq q_l$ for any $l$, we noted that $l_{i+1}=l_i$. Thus, the factors in the product coming from these terms cancel out and only the terms in $q_l$ remain. The claim is proven for $n=1$.

If $n>1$, the same procedure applies. For example, one could pick $\{t^{(m)}_j\}_j $ where $m$ indexes the points from $1$ to $n$. Then, one considers $\{r_i\}=\{q_l\}\bigcup_m\{t^{(m)}_j\}$ as before. In this case, at each level $l$, every equivalence class $\bar{i}$ of $\{1,...,n\}/\sim_l$ picks up a Radon-Nikodym derivative $V_{\bar{i}}$. 
\end{proof}
We now state a result on the regularity of the expectations of the field $\kappat$ that we will need when studying functionals of the GREM's. We omit the proof as it is a direct consequence of the fact that $\tilde{\mu}^{(n)}_Q$ is smooth in the parameters $x_l$'s and $q_l$'s.
\begin{cor}[Expectations of the Backward Field]
Let $x\in\maone$. Consider $(\xi,Q)$, a GREM($x$), with a gaussian field $\kappa$. Let $Q_n$ be an $n\times n$ matrix such that $\prob_x(Q_n)\neq 0$ and define
$$ F_{Q_n}\left(\mathbf{x},\mathbf{q}\right)= \E_x\left[\prod_{i=1}^n\phi(\kappat_i(1))\Big| Q_n \right]$$
where $\phi:\R\to\R$ is a bounded continuous function and $\mathbf{q}=(q_0,...,q_k)$ and  $\mathbf{x}=(x_0,...,x_k)$ as before. The following statements hold :
\begin{enumerate}[i]
\item $F_{Q_n}\left(\mathbf{x},\mathbf{q}\right)$ is a continuous function of $\mathbf{q}$;
\item $F_{Q_n}\left(\mathbf{x},\mathbf{q}\right)$  is continuous in $x_j$ on $(x_{j-1},x_{j+1})$, for all $1\leq j\leq k$. The limits $\lim_{x_j\to x_{j-1}}F_{Q_n}(\mathbf{x},\mathbf{q})$ and $\lim_{x_j\to x_{j+1}}F_{Q_n}(\mathbf{x},\mathbf{q})$ exists and are continuous functions of $\mathbf{q}$.
\end{enumerate}
\label{continuity}
\end{cor}
In the last claim, we stress out that the limit of $F_{Q_n}$ as $x_j\to x_{j+1}$ is not equal to $F_{Q_n}$ evaluated at $(x_0,...,x_{j-1},x_{j+1},x_{j+1},...,x_k)$ and similarly for the limit $x_j\to x_{j-1}$. This is basically because the distribution functions of the approximating sequence clearly possess one atom more than the limiting distribution function with $x_j=x_{j+1}$. Therefore, $\lim_{x_j\to x_{j+1}}  F_{Q_n}\left(\mathbf{x},\mathbf{q}\right)$ differs from the function $F_{Q_n}$ evaluated at the limiting distribution function as the product over the number of atoms appearing in the density of $\kappat$ in Proposition \ref{backward cavity} has an extra factor in the first case.

\section{The Parisi Functional as a functional on RPC's}
In this section, we establish that functionals on RPC's of the form \eqref{intro fct} coincide with Parisi-like functionals on the space $\maone$. The RPC formulation is a natural framework for deriving properties of such functionals as one can take advantage of the rich structure of the GREM and of the quasi-stationarity property. This is useful as it can be sometimes tedious to derive properties from the solution to the differential equation \eqref{intro diff eqn}. As an example of techniques in the RPC formulation, we obtain differentiation formulas which lead to the continuity and the monotonicity of the functional. This result slightly generalizes a theorem of Guerra \cite{Guerra_bound}. Using continuity, we also exhibit a limit form of the RPC functional in the singular case $x\in\ma$ but $x\notin\maone$.

\subsection{The Parisi Functional }
We first define the Parisi functional in a general way. The reader can consult \cite{Guerra_bound, Panchenko, Talagrand_parisi} for particular choices of settings.

\begin{df}
Let $\psi\in \C$ and $g\in C^1([0,1])$, a strictly increasing function such that $g(0)=0$. The Parisi functional $\Ppargen^{par}:\ma\to \R$ is defined as $\Ppargen^{par}(x):=f_x(0,0)$
where $f_x(q,y)$ is the solution to the differential equation
\begin{equation*}
\partial_q f(q,y)+\frac{g'(q)}{2}\left[\partial^2_yf(q,y)+x(q) \left(\partial_y f(q,y)\right)^2\right]=0
\end{equation*} 
with the boundary condition $f(1,y)=\psi(y)$.  The case $\psi(\eta)=\log\cosh(\beta\eta+h)$ and $g(q)=q$ reduces to equation \eqref{intro diff eqn}.
\end{df}

From another perspective, the $AS^2$ variational principle formulation is based on functionals of the following form \cite{ASS,ASS2}.
\begin{df}
Let $\psi$ and $g$ be as above. The functional $\Ppargen:\maone\to\R$ is defined by
\begin{equation}
\Ppargen(x):=\E_x\left[\log \frac{\sum_{\alpha}\sa e^{\psi(\kappa_\alpha(1))}}{\sum_{\alpha}\sa}\right]
\label{ruelle_functional}
\end{equation}
where $(\xi,Q)$ is a GREM($x$) and $\kappa$ is a gaussian field on $\xi$ with covariance $ \text{Cov}(\kappa_\alpha(q),\kappa_{\alpha'}(q'))= g(q\wedge q'\wedge q_{\alpha,\alpha'})$.
We will write $\Ppar$ in the special case $\psi(\eta)=\log\cosh(\beta\eta+h)$ and $g(q)=q$.
\end{df}
The main result of the section is the correspondence of these functionals and the Parisi functionals.
\begin{thm}[The Parisi functional on RPC's]

For any $x\in\maone$, $\Ppargen^{par}(x)=\Ppargen(x).$
\label{ruelle=parisi}
\end{thm}
\begin{proof}
The quasi-stationarity of the GREM (Theorem \ref{GREM QS}) implied equation \eqref{QS fct}:
$$\xi_{\alpha}e^{\psi(\kappa_\alpha(1))}\distrib \xi_\alpha e^{\psi_{q_1}(\kappa_{\alpha(0)}(q_1))}.$$
By inserting this into equation \eqref{ruelle_functional}, one gets
$$\Ppargen(x)=\E_x\left[\psi_{q_1}(\kappa_{\alpha(0)}(q_1))\right]=\psi_0(0)$$
by definition of $\psi_0$ (equation \eqref{psiq}). 
But $\psi_q(y)$ is the solution to the differential equation of the Parisi functional by Theorem \ref{GREM QS} again. We conclude that $\Ppargen(x)=\Ppargen^{par}(x)$. 
\end{proof}

It is worth pointing out that $\Ppargen$ is linear when $\psi$ is.
\begin{prop}[Linearity of $\Ppargen$]
If $\psi(\eta)=\beta\eta$, then $ \Ppargen(x)$ is a bounded linear functional. Precisely, $\Ppargen(x)=\frac{\beta^2}{2}\int_0^1x(q)dg(q)$.
\label{fct_lin}
\end{prop}
\begin{proof}
We can write $\kappa_\alpha(1)$ as a sum of independent increments: $\kappa_{\alpha}(1)=\sum_{l=0}^{k}\delta_{\alpha(l)}$ where $\delta_{\alpha(l)}=\kappa_\alpha(q_{l+1})-\kappa_\alpha(q_{l})$. We can then apply the quasi-stationarity to each REM $\zeta^{\alpha(l-1)}$ of the cascade:
$$ \xi_\alpha e^{\beta\sum_{l=0}^{k}\delta_{\alpha(l)}}\distrib e^{\beta\kappa_{\alpha(0)}(q_1)}\xi_\alpha \prod_{l=1}^ke^{\frac{\beta^2}{2}x_l(g(q_{l+1})-g(q_{l}))}$$
where we have used the fact that the Laplace transform of a standard gaussian is $e^{\frac{\lambda^2}{2}}$. Note that the expectation of $\kappa_{\alpha(0)}(q_1)$ is $0$ by definition. 
The claim is proven by inserting the last equation into equation \eqref{ruelle_functional} for $\Ppargen$.
\end{proof}

\subsection{Differentiation}
How does $\Ppargen$ vary as the position $q_l$ of an atom or the weight $x(q_l)$ is changed? To address this question, we look at the derivatives of $\Ppargen(x)$ with respect to the parameters $x_l$ and $q_l$. It turns out that the $x$-derivatives are simply related to the $q$-derivatives due to the structure of the space $\ma$. Both are expressed as expectations on the backward field $\kappat$.
 For conciseness, we will use the notation $\kappa_i\equiv\kappa_i(1)$ for a gaussian field $\kappa_i$ at $q=1$ throughout the section.
\begin{prop}[$q$-Derivatives]
Let $x\in\maone$ with atoms $\{q_l\}$. 
The following differentiation formulas hold:
\begin{equation}
\partial_{q_l}\Ppargen(x)=\frac{-g'(q_l)}{2}\E_x\left[\psi'(\kappat_1)\psi'(\kappat_2)\chi_{q_{12}=q_l}\right]
\label{1st deriv q}
\end{equation}
 where $\kappa$ is a gaussian field on a GREM($x$) and $\id_A$ is the identity function of the event $A$. If $g(q)=q$,
%\begin{equation}
\begin{align}
\partial_{q_i}\partial_{q_j}\Ppargen(x)&=-\frac{3}{2}\E_x\left[\prod_{m=1}^4\psi'(\etat_m)\id_{q_{12}=q_i,q_{34}=q_j}\right]
\label{2nd deriv q} \\
&+2\E_x\left[\psi'(\etat_1)\left(\psi''(\etat_2)+\psi'^2(\etat_2)\right)\psi'(\etat_3)\id_{q_{12}=q_i,q_{23}=q_j}\right]\nonumber\\
&-\frac{1}{2}\delta_{ij}\E_x\left[\left(\psi''(\etat_1)+\psi'^2(\etat_1)\right)\left(\psi''(\etat_2)+\psi'^2(\etat_2)\right)\id_{q_{12}=q_i}\right]\nonumber
\end{align}
where $\eta$ is the cavity field of the GREM($x$).
In particular, both derivatives are continuous functions of the $q_l$'s.
%\end{equation}
\label{q-diff Ppar}
\end{prop}
\begin{proof}
The continuity of the derivatives follows from the form of the expression and Corollary \ref{continuity}.  
For the first derivative, we use the gaussian differentiation formula \eqref{wick 1st deriv} in Appendix \ref{app:gaussian}. The parameter is $q_l$ and the covariance is $\E_x[\kappa_\alpha(1)\kappa_{\alpha'}(1)]=\sum_{l=1}^{k+1} g(q_l)\delta_{\qaa,q_l}$ where $\delta_{q,q'}=1$ if $q=q'$ and $0$ otherwise. Therefore $\partial_{q_l}\E_x[\kappa_\alpha(1)\kappa_{\alpha'}(1)]=g'(q_l)\delta_{\qaa=q_l}$
for $l=1,...,k$. Note that the first term of formula \eqref{wick 1st deriv} vanishes as $q_{\alpha,\alpha}=1$ for all $\alpha$. Thus, the first derivative becomes
$$  \partial_{q_l}\Ppargen(x)=\frac{-g'(q_l)}{2}\E_x\left[\frac{\sum_{\alpha,\alpha'}\sat\tilde{\xi}_{\alpha'}\psi'(\kappa_\alpha)\psi'(\kappa_{\alpha'})\delta_{\qaa,q_l}}{\sum_{\alpha,\alpha'}\sat\tilde{\xi}_{\alpha'}}\right]$$
where we have used the notation $\sat=\sa e^{\psi(\kappa_\alpha(1))}$. After ordering the $\sat$, we can write  by the definition of $\kappat$
$$ \partial_{q_l}\Ppargen(x)=\frac{-g'(q_l)}{2}\E_x\left[\frac{\sum_{i,j}\xit_{i}\xit_{j}\psi'(\kappat_i)\psi'(\kappat_{j})\delta_{q_{ij},q_l}}{\sum_{i,j}\xit_i\xit_j}\right].$$
Conditioning on $\xit$ and $q_{ij}$, bearing in mind the independence of $\kappat $ and $\xit$ from Proposition \ref{backward cavity}, one can write the expectation in the r.h.s. as
$$\E_x\left[\psi'(\kappat_1)\psi'(\kappat_2)|q_{12}=q_l\right] \E_x\left[\frac{\sum_{i,j}\xit_{i}\xit_{j}\delta_{q_{ij},q_l}}{\sum_{i,j}\xit_i\xit_j}\right].$$
where we have used the fact that the joint distribution of $\kappat_i$ and $\kappat_j$ is the same as the distribution of $\kappat_1$ and $\kappat_2$ for all $i,j$. The proof is completed by first recalling that the shifted normalized process $\xit_i/\sum_i\xit_i$ is distributed as the original normalized process by the quasi-stationarity property. Moreover, the normalized process is independent of the overlap matrix $\{q_{ij}\}$ as it was remarked in Theorem \ref{prob_grem}. Therefore, the second expectation above simply becomes $\prob_x(q_{12}=q_l)$ by conditioning on the normalized weights and by noticing that $\prob_x(q_{ij}=q_l)$ does not depend on $i,j$ for $i\neq j$. This yields equation \eqref{1st deriv q}.

The second expression is obtained from formula \eqref{wick 2nd deriv} in Appendix \ref{app:gaussian} with $\phi(\kappa)=\log\sum_\alpha\sa e^{\psi(\kappa_\alpha)}$ and straightforward derivation.
\end{proof}

Obviously, a formula holds also for a more general $g$ in the case of the second derivatives. We omit it for the sake of conciseness.

The nice feature of the above derivative formulas is that they explicitly express the derivatives of the Parisi functional in terms of expectations over the field $\kappat$ and probabilities on the GREM cascade. Using conditional expectations, we may rewrite the second derivative formula, e.g. in the case $\psi(z)=\log\cosh(\beta z+h)$ where the identity $\psi''+\psi'^2=1$ holds, as 
\begin{align}
\partial_{q_i}\partial_{q_j}\Ppar(x)=&-\frac{3}{2}\beta^4\sum_{Q_4(q_i,q_j)}\prob_x(Q_4(q_i,q_j))\E_x\left[\prod_{m=1}^4\tanh(\beta \etat_m+h)\Big|Q_4(q_i,q_j)\right]
\label{2nd deriv q SK} \\
&+2\beta^4\sum_{Q_3(q_i,q_j)}\prob_x(Q_3(q_i,q_j))\E_x\left[\tanh(\beta\etat_1+h)\tanh(\beta\etat_2+h)|Q_3(q_i,q_j)\right]\nonumber\\
&-\frac{1}{2}\beta^4\delta_{ij}\prob_x(q_{12}=q_i)\nonumber.
\end{align}
where the sums are over all $3\times 3$ and $4\times 4$ matrices $Q_3$ and $Q_4$ such that $q_{12}=q_i,q_{23}=q_j$ and $q_{12}=q_i,q_{34}=q_j$ with $\prob_x(Q_3(q_i,q_j))\neq 0$ and $\prob_x(Q_4(q_i,q_j))\neq 0$. Note in particular that these matrices must satisfy the inequality \eqref{um3}.

\begin{cor}[Useful estimate on the $q$-derivatives]
Let $x\in\maone$ with $i$-th atom at $q_i$. Let $C>0$ be such that $|\psi'|\leq C$. Then
\begin{equation}
0\leq -\partial_{q_i}\Ppargen(x)\leq \frac{C^2}{2}(x_i-x_{i-1})g'(q_i).
\end{equation}
\label{positivity q}
\end{cor}
\begin{proof}
The upper bound is clear from equation \eqref{1st deriv q}, the assumption on $\psi$ and the fact that $\prob_x(q_{12}=q_i)=x_i-x_{i-1}$. The lower bound is a consequence of
$$\E_x\left[\psi'(\etat_1)\psi'(\etat_2)\Big|q_{12}=q_i\right]=\E_x\left[\E_x[\psi'(\etat_1)|\F_{q_i}]^2\right] $$
\end{proof}

The structure of the space $\ma$ enables us to relate the $x$-derivatives and the $q$-derivatives through a simple differentiation scheme. 
To do so, we need to define the measure obtained from $x\in\ma$ by transporting a mass $\delta$ from an atom at $q_i$ to $s\in[0,1]$.
\begin{df}
Let $x\in \ma$. Let $q_i$ be the position of the $i$-th atom of $x$, $i\leq k$. Consider $\delta>0$ and $s\in[0,1]$. We define $x_{\delta,q_i\to s}\in\ma$ as the atomic measure obtained from $x$ by transporting a mass $\delta$ from $q_i$ to $s$.
Note that if $s=q_j$ for some $j$, then $x_{\delta,q_i\to s}$ has $k$ atoms; otherwise, $x_{\delta,q_i\to s}$ has $k+1$ atoms.
\label{def mass transport}
\end{df}

\begin{prop}[$x$-Derivatives]
Consider $x\in\maone$. Then
\begin{equation}
\partial_{x_i}\Ppargen(x)=\frac{1}{2}\int_{q_i}^{q_{i+1}}\lim_{\delta\to 0} \E_{x_{\delta,q_{i+1}\to r}}
\left[\psi'(\kappat_1)\psi'(\kappat_2)\Big|q_{12}=r\right]dg(r).
\label{1st deriv x}
\end{equation}
In particular, if $|\psi'|\leq C$ for some $C>0$
$$0\leq \partial_{x_i}\Ppargen(x)\leq \frac{C^2}{2}(g(q_{i+1})-g(q_i)).$$
 \label{x-diff Ppar}
\end{prop}
%Note that by, Lemma \ref{continuity in delta}, the limiting function is well-defined.

\begin{proof}
The upper and lower bounds are obtained as in Corollary \ref{positivity q} followed by integration. 

For the sake of clarity, we set $x_\delta=x_{\delta,q_{i+1}\to r}$ throughout the proof.
By Proposition \ref{q-diff Ppar},
$$-\partial_r\Ppargen(x_\delta)= \delta \frac{1}{2}g'(r)\E_{x_\delta}\left[\prod_{m=1}^2\psi'(\kappat_m)\Big|q_{12}=r\right].$$
Here we have used $\prob_{x_\delta}(q_{12}=r)=x_\delta(r)-x_\delta(q_i)=\delta$. The r.h.s. of the above equation satisfies the fundamental theorem of calculus as a function of $r$ as $g'(r)$ and $\E_{x_\delta}\left[\prod_{i=1}^2\psi'(\kappat_i)\Big|q_{12}=r\right]$ are continuous functions of $r$ by definition and Corollary \ref{continuity} respectively.
Moreover, the limit $\lim_{\delta\to 0}\frac{1}{\delta}\partial_r\Ppargen(x_{\delta})$ exists and is bounded thanks to Lemma \ref{continuity}:
$$ \lim_{\delta\to 0}\frac{1}{\delta}\partial_r\Ppargen(x_{\delta})=\frac{1}{2}g'(r)\lim_{\delta\to 0}\E_{x_\delta}\left[\prod_{m=1}^2\psi'(\kappat_m)\Big|q_{12}=r\right].$$
The fundamental theorem of calculus and the above limit yield the desired expression for the right-derivative $\partial^+_{x_i}\Ppargen(x)$:
\begin{align*}
\partial^+_{x_i}\Ppargen(x)&=\lim_{\delta\to 0}\frac{\Ppargen(x_{\delta,q_{i+1}\to q_i})-\Ppargen(x)}{\delta}\\
&= \lim_{\delta\to 0}\int_{q_i}^{q_{i+1}}\frac{-\partial_r\Ppargen(x_\delta)}{\delta}\\
&=  \int_{q_i}^{q_{i+1}}\lim_{\delta\to 0}\frac{-\partial_r\Ppargen(x_\delta)}{\delta}.
\end{align*}
The equality of left and right derivatives is checked using the fact that $\lim_{\delta\to 0}x_{\delta,q_{i+1}\to r}=\lim_{\delta\to 0}x_{\delta,q_{i}\to r}$ and the $x$-continuity in Corollary \ref{continuity}.
\end{proof}
The simplicity of the differentiation scheme gives a formal approach for computing derivatives of all orders in $x$ for RPC's functionals. 

\subsection{Continuity and Monotonicity}
As a direct application of the differentiation formulas, we establish the $L^1$-continuity and the monotonicity of $\Ppargen$. This is a theorem due to Guerra \cite{Guerra_bound} that we prove in a general setting.

The space $\mc$ has a natural partial ordering. We say that $x\in\mc$ {\it dominates} $y\in\mc$ if $x(q)\geq y(q)$ for all $q\in[0,1]$. The terminology refers to the stochastic dominance of the random variables associated to the distribution functions. A functional $\Lambda:\mc\to\R$ is said to be {\it monotone increasing} with respect to this partial ordering if for any $x,y\in\mc$ such that $x$ dominates $y$, $\Lambda(x)\geq \Lambda(y)$. 

We start by stating a useful lemma whose proof is straightforward.
\begin{lem}
Let $x,y\in\ma$ and $\Lambda:\ma\to\R$. Suppose $\partial_{x_i}\Lambda$ exists for all $x\in\ma$.  Suppose also that the estimate $ 0\leq\partial_{x_i}\Lambda(x)\leq C_\Lambda (g(q_{i+1})-g(q_i))$ holds for some $C_\Lambda>0$ and for a strictly increasing function $g\in C^1([0,1])$ such that $g(0)=0$. Then
\begin{equation}
\Lambda(x)-\Lambda(y)\leq C_\Lambda\int_0^1 \max\{x(q)-y(q),0\} dg(q).
\label{estimate}
\end{equation}
\label{lem_continuity}
\end{lem}
The theorem of Guerra appeared in \cite{Guerra_bound} without proofs. 
A proof can be found in \cite{Talagrand_parisi}. In the RPC formulation, it is a basic consequence of the bound on the $x$-derivative.
\begin{thm}[Continuity and Mononicity]
Let $\psi\in\C$ such that $|\psi'|\leq C$ and $g\in C^1([0,1])$, a strictly increasing function with $g(0)=0$. Then the following holds:
\begin{enumerate}
\item If $x,y\in\maone$, 
\begin{equation*}
|\Ppargen(x)-\Ppargen(y)|\leq \frac{C^2}{2} \|x-y\|_{L^1(g'(q)dq)}.
\end{equation*}
In particular, $\Ppargen$ has a continuous extension to the whole set $\mc$ in the $L^1$-topology (and so in the weak topology).
\item $\Ppargen$ is monotone increasing on $\mc$. 
\end{enumerate}
\label{guerra_thm}
\end{thm}
\begin{proof}
The bounds of Proposition \ref{x-diff Ppar} satisfy the assumptions of Lemma \ref{lem_continuity}. The estimate of the first claim follows from equation \eqref{estimate}. The $L^1$-continuity is clear from the estimate. The continuous extension is possible as $\maone$ is dense subset of $\mc$ on the $L^1$-topology. The monotonicity is proven by Lemma \ref{lem_continuity} and the fact that if $x,y\in\ma$ and $x(q)\geq y(q)$ for every $q\in[0,1]$, then $\max\{0,y(q)-x(q)\}=0$. The property is extended to $\mc$ by continuity.
\end{proof}

\subsection{The singular case $x=1$}
When introducing the REM, we noticed that the case $x=1$ was singular as far as the intensity measure is concerned. However, it is necessary in the SK model theory to consider functionals on GREM's whose last level of splitting consist formally of REM($1$). These GREM's correspond to elements of $\ma$ not in $\maone$. Using $L^1$-continuity, we are able to obtain an expression for $\Ppargen$ and its differentiation formulas \eqref{1st deriv q}, \eqref{2nd deriv q} and \eqref{1st deriv x} evaluated at these singular GREM's. 

Let $x\in\ma$ with $k$ atoms at $\{q_l\}$ and $q_k<1$. Then $x\notin\maone$. Consider $x^\epsilon\in\maone$ with $x^\epsilon(q_l)=x(q_l)$ for $1\leq l< k$, $x^\epsilon(q_k)=1-\epsilon$ and $x^\epsilon(q_{k+1})=1$. Clearly, $x^\epsilon\to x$ in $L^1(g'(q)dq)$ as $\epsilon$ goes to $0$. Applying quasi-stationarity to a REM($1-\epsilon$) with shift $e^{ \psi(\kappa_\alpha(1))}$, we pick up the factor $\E_{x^\epsilon}\left[e^{(1-\epsilon) \psi(\kappa_\alpha(1))}|\F_{q_k}\right]^{1/1-\epsilon}$
which simply becomes in the limit : $\E_{x^\epsilon}\left[e^{ \psi(\kappa_\alpha(1))}|\F_{q_k}\right]=:e^{\psi_{q_k}(\kappa_{\alpha(k-1)}(q_k))}$. One applies $L^1$-continuity and the dominated convergence theorem to $\Ppargen(x^\epsilon)$ to get
$$\Ppargen(x)=\E_x\left[\log\frac{\sum_{\alpha(k-1)}\xi_{\alpha(k-1)}e^{\psi_{q_k}(\kappa_{\alpha(k-1)}(q_k))}}{\sum_{\alpha(k-1)}\xi_{\alpha(k-1)}}\right]. $$

The cases $\psi(\eta)=\log\cosh(\beta\eta+h)$ and $\psi(\eta)=\beta\eta$ are again special as
$$e^{\psi_{q_k}(\kappa_{\alpha(k-1)}(q_k))}:= E_z\left[e^{ \psi(z\sqrt{g(1)-g(q_k)}+\kappa_{\alpha(k-1)}(q_k))}\right]=e^{\frac{\beta^2}{2}(g(1)-g(q))}e^{\psi(\kappa_{\alpha(k-1)}(q_k))} $$
and $\psi$ is retrieved after integration. In these cases,
\begin{equation}
\Ppargen(x)=\frac{\beta^2}{2}(g(1)-g(q))+\E_x\left[\log\frac{\sum_{\alpha(k-1)}\xi_{\alpha(k-1)}e^{\psi(\kappa_{\alpha(k-1)}(q_k))}}{\sum_{\alpha(k-1)}\xi_{\alpha(k-1)}}\right].
\label{ppar ma}
\end{equation}

In particular, if $k=1$, i.e. $x$ has a single atom sitting at $q$:
\begin{equation}
\Ppargen(x)=\frac{\beta^2}{2}(g(1)-g(q))+\E_x\left[\psi(\kappa_{\alpha(0)}(q))\right]=\frac{\beta^2}{2}(g(1)-g(q))+\int_\R \frac{e^{-\frac{z^2}{2}}}{\sqrt{2\pi}}\psi(z\sqrt{g(q)})dz
\label{ht 1}
\end{equation}
which, not surprisingly, resembles the high-temperature solution of the SK model.

The derivative formulas \eqref{1st deriv q}, \eqref{2nd deriv q} and \eqref{1st deriv x} are retrieved by applying the derivative to the expression \eqref{ppar ma}. The formulas hold by replacing $\kappat_i(1)$ by $\kappat_i(q_k)$. As an example, we compute $\partial_{q_k}\Ppar(x)$:
\begin{align}
\partial_{q_k}\Ppar(x)&=-g'(q_k)\frac{\beta^2}{2}+g'(q_k)\frac{\beta^2}{2}\E_x\left[(\psi''+\psi'^2)(\kappat_{1}(q_k))\right]\nonumber\\
&-g'(q_k)\frac{\beta^2}{2}\E_x\left[\psi'(\kappat_{1}(q_k))\psi'(\kappat_{2}(q_k))\chi_{q_{12}=q_k}\right]\nonumber\\
&=-g'(q_k)\frac{\beta^2}{2}\E_x\left[(\psi')^2(\kappat_{1}(q_k))\chi_{q_{12}=q_k}\right]\label{diff k=1}
\end{align}
where we used the differentiation formula \eqref{wick 1st deriv} in the first equality and the identity $\psi''+\psi'^2=1$ in the second.

\section{The $AS^2$-Variational Principle and the Parisi Formula}
The $AS^2$ variational principle expresses the pressure of the SK model in the thermodynamic limit $P_{SK}(\beta,h)$ as an optimization problem over random overlap structures \cite{ASS, ASS2}:
\begin{equation}
P_{SK}(\beta,h)=\lim_{M\to\infty}\inf_{\{\mu \text{ ROSt}\}}G_M(\beta,h,\mu)
\label{var princ}
\end{equation}
where $\mu$ is the probability measure of a ROSt $(\xi,Q)$. The functional $G_M$ is given by
\begin{equation*}
G_M(\beta,h,\mu):=\frac{1}{M}\E_\mu\left[\log\frac{\sum_{\alpha}\sa \prod_{i=1}^M2\cosh(\beta\eta^i_\alpha+h)}{\sum_{\alpha}\sa \prod_{i=1}^M e^{\beta \kappa^i_\alpha}}\right]
\end{equation*}
where $\eta^i$ and $\kappa^i$, $i=1,...,M$, are independent copies of the cavity field and of a gaussian field with covariance function $q^2/2$.

As we noticed previously, the RPC's form a particularly interesting class of ROSt's due to the ultrametric structure of the overlap matrix and the quasi-stationarity property. In this section, we study the $AS^2$ variational principle \eqref{var princ} restricted to the class of RPC's. First, we show that the restricted variational problem reduces to the Parisi formula. Then, we rederive the Almeida-Thouless line which yields sufficient condition for the minimizer of the variational problem not to be a single atom. For this, we follow the idea of Toninelli but we explicitly use the RPC structure underlying the functionals.

\subsection{The Parisi Formula}
The Parisi formula for the pressure of the SK model was proven by Talagrand in \cite{zeproof}.
\begin{thm}[The Parisi Formula]
\begin{equation*}
P_{SK}(\beta,h)=\inf_{x\in \ma} \left\{\log 2+\Ppar^{par}(x)-\frac{\beta^2}{2}\int_0^1qx(q)dq\right\}.
\label{Parisi_formula}
\end{equation*}
\end{thm}
It is remarked in \cite{ASS, ASS2} that the Parisi formula is exactly the $AS^2$ variational problem restricted to the class of RPC's. To establish this connection, we start by noting that the limit $M\to\infty$ in equation \eqref{var princ} is no longer needed when we deal with RPC's.
\begin{prop}[Variational Principle over the class of GREM's]
Let $\mu$ be a GREM parametrized by $x\in\maone$. Then, for any $M\in\N$, $G_M(\beta,h,\mu)=G_1(\beta,h,\mu)$ and
$$\Gpar(x):=G_1(\beta,h,\mu)=\E_x\left[\log \frac{\sum_{\alpha}\sa 2\cosh(\beta\eta_\alpha(1)+h)}{\sum_{\alpha}\sa}\right] -\E_x\left[\log \frac{\sum_{\alpha}\sa e^{\beta\kappa_\alpha(1)}}{\sum_{\alpha}\sa}\right].$$
The variational problem of equation \eqref{var princ} restricted to GREM's reduces to
\begin{equation*}
\lim_{M\to\infty}\inf_{\{\mu \text{ GREM}\}}G_M(\beta,h,\mu)=\inf_{x\in\ma}\Gpar(x).
\end{equation*}
\label{G for GREM}
\end{prop}
\begin{proof}
It suffices to note that each of the $M$ independent copy of the fields contributes the same factor to the pressure. This is done using equation \eqref{QS fct}. 
\end{proof}
Note that $G_{\beta,h}(x)$ is the difference of two RPC functionals with $\psi(\eta)=\log\cosh(\beta\eta+h)$, $g(q)=q$ and $\psi(\kappa)=\beta\kappa$, $g(q)=q^2/2$ respectively. The Parisi formula is retrieved from the $AS^2$ variational principle on RPC's by using Theorem \ref{ruelle=parisi} and Proposition \ref{fct_lin} for these two functionals:
\begin{prop}[The Parisi Formula with RPC's]
\begin{equation}
P_{SK}(\beta,h)=\inf_{\text{GREM($x$)}} G_{\beta,h}(x).
\label{Parisi_formula2}
\end{equation}
\end{prop}
As $\Gpar$ is the difference of two $L^1$-continuous functionals by Theorem \ref{guerra_thm}, it is itself $L^1$-continuous and can be extended continuously to $\mc$. Therefore we can rewrite equation \eqref{Parisi_formula2} as
\begin{equation}
 P_{SK}(\beta,h)=\min_{x\in \mc} \Gpar(x)
 \label{Parisi_formula3}
\end{equation}
because $\mc$ is compact in the $L^1$-topology (recall that the $L^1$-topology is equivalent to the weak topology. See Appendix \ref{app:topo}).
The question of the uniqueness of the minimizer of the Parisi formula was raised in \cite{Panchenko, Talagrand_parisi} but remains open.

\subsection{An example of calculation: The Almeida-Thouless Line}
It is now well known that the minimizer of the variational principle \eqref{Parisi_formula3} for the SK model is an atomic measure with a single atom when $\beta$, as a function of $h$, is small enough \cite{ALR, GT_quadratic}. This is referred to as the high-temperature solution of the SK model. Moreover, it was proven by Toninelli that this solution cannot hold beyond the so-called Almeida-Thouless line \cite{AT}. In this section, we rederive this sufficient condition following Toninelli in spirit, but relying heavily on the RPC structure. In doing so, we hope to illustrate the convenient features of the RPC formalism\footnote{It was recently proven by Guerra that the high-temperature solution actually holds up to the Almeida-Thouless line. This had been rigorously established only in the case $h=0$ \cite{Guerra_AT}.}.

First, we remark that the stationarity conditions of the optimization problem \eqref{Parisi_formula3} $\partial_{q_i} \Gpar(x)=0$ for $i=0,1,...,k$ yield self-consistency equations:
\begin{equation*}
\E_x\left[\tanh(\beta\etat_1(1)+h)\tanh(\beta\etat_2(1)+h)|q_{12}=q_i\right]=q_i.
\end{equation*}
This is a consequence of the definition of $\Gpar$ and Proposition \ref{q-diff Ppar}.
In the case of a single atom, we get the self-consistency equation for the high-temperature solution from equation \eqref{diff k=1}
\begin{equation*}
E_z\left[\tanh^2(\beta z\sqrt{q}+h)\right]=q.
\end{equation*}
We denote a solution to the self-consistency equation $\qbar=\qbar(\beta,h)$. This solution is unique in the case $\beta<1$ and $h=0$ and in the case $h\neq 0$ (see e.g. \cite{GT_quadratic}). 

Fix $\beta>0$ and $h\in\R$. The high-temperature solution corresponds to the infimum of $\Gpar$ over the subset of $\ma$ consisting of measures with a single atom. Note that this set is compact so the infimum is attained. Let $x^*\equiv x^*(\beta,h)$ be the minimizer. The only atom of $x^*$ must be located at $\qbar$, the solution to the self-consistency equation. The idea for deriving the the Almeida-Thouless condition is to show that, if $\beta$ and $h$ are such that
$$\beta^2\int_\R \frac{e^{-z^2/2}}{\sqrt{2\pi}}\cosh^{-4}(\beta z\sqrt{\qbar}+h)> 1,$$
then there exists an element of $\ma$ with two atoms such that $\Gpar$ evaluated at that element is smaller than $\Gpar(x^*)$. This implies that the high-temperature solution cannot hold for the optimization problem \eqref{Parisi_formula3} in this region of the plane $(\beta,h)$.

Let us construct such an element. Pick $0\leq m\leq 1$ and $\qbar\leq r\leq 1$. Let $x^*_{m,r}$ be the atomic measures with atoms at $\qbar$ and $r$ with $x^*_{m,r}(\qbar)=m$ and $x^*_{m,r}(r)=1$. Note that $x^*_{m,\qbar}=x^*$. As $\Gpar(x^*_{m,r})$ is continuously differentiable in $m$ on $0<m<1$ for $r>\qbar$, we have
\begin{equation}
 \Gpar(x^*)=\Gpar(x^*_{m,r})+\int_{m}^1\partial_{m'}\Gpar(x^*_{m',r})dm'.
 \label{integral formula}
 \end{equation}
From equation \eqref{integral formula}, we see that $\Gpar(x^*_{m,r})<\Gpar(x^*)$ if $r$ and $m$ are such that
\begin{equation}
\partial_{m'}\Gpar(x^*_{m',r}) >0
\label{to prove AT}
\end{equation}
for $m\leq m' <1$.
To get this inequality, we follow \cite{AT} and expand $\partial_{m}\Gpar(x^*_{m,r})$ around $\qbar$
\begin{align}
\partial_{m}\Gpar(x^*_{m,r})&=\partial_{m}\Gpar(x^*_{m,r})\eval_{r=\qbar}+ (r-\qbar)\partial_r\partial_{m}\Gpar(x^*_{m,r})\eval_{r=\qbar}\label{taylor}\\
&+\frac{(r-\qbar)^2}{2}\partial^2_r\partial_{m}\Gpar(x^*_{m,r})\eval_{r=\qbar} +\mathcal{O}\left((r-\qbar)^3\right).\nonumber
\end{align}
The remainder term is bounded (this can be checked using the gaussian differentiation formula to calculate the third derivative).

The first term of the expansion is $0$ by Proposition \ref{x-diff Ppar} as the integral involved in the differentiation formula is from $\qbar$ to $r$. The second term vanishes too as we retrieve the self-consistency equation for $\qbar$ by differentiating in $r$ the integral expression for $\partial_{m}\Gpar(x^*_{m,r})$
$$\partial_r\partial_{m}\Gpar(x^*_{m,r})\eval_{r=\qbar}=\E_{x^*_{m,r}}\left[\tanh(\beta\etat_1(r)+h)\tanh(\beta\etat_2(r)+h)|q_{12}=q_i\right]-r\eval_{r=\qbar}=0.$$
Therefore to prove \eqref{to prove AT}, it suffices to find conditions for which $\partial_{m}\partial^2_r\Gpar(x^*_{m,r})\eval_{r=\qbar}>0. $
Equation \eqref{2nd deriv q SK} is useful to compute $\partial^2_r\Gpar(x^*_{m,r})$. From Theorem \ref{prob_grem}, the matrices $Q_3$ and $Q_4$ with $\prob_{x^*_{m,r}}(Q_3)\neq 0$ and $\prob_{x^*_{m,r}}(Q_4)\neq0$ are
\begin{align}
\prob_{x^*_{m,r}}(q_{12}=r)&=1-m\nonumber\\
\prob_{x^*_{m,r}}(q_{12}=r,q_{23}=r)&=\frac{(2-m)(1-m)}{2}\nonumber\\
\prob_{x^*_{m,r}}(q_{12}=r,q_{34}=r, q_{13}=r)&= \frac{(3-m)(2-m)(1-m)}{6}\nonumber\\
\prob_{x^*_{m,r}}(q_{12}=r,q_{34}=r,q_{13}=\qbar)&=\frac{m(1-m)^2}{6}.\nonumber
\end{align}
The reader can check that all missing overlaps $q_{ij}$ of the matrices $Q_3$ and $Q_4$ in the above events are determined by ultrametricity. For example, if  $q_{12}=r,q_{23}=r$ then $q_{13}=r$.
From the probabilities above, one can see that, when applying the derivative $\partial_m$ directly to $\partial^2_r\Gpar(x^*_{m,r})$ and taking the limit $m\to1^-$, only the terms coming from the derivative of the factor $1-m$ do not vanish. Thus, one gets the remaining terms
\begin{align}
& \lim_{m\to1^-}\left(\partial^2_r\partial_{m}\Gpar(x^*_{m,r})\eval_{r=\qbar}\right)=
-\frac{\beta^2}{2}\left(1-\beta^2\E_{x^*}\left[1-2\tanh^2(\beta(\etat(\qbar)+h))+\tanh^4(\beta(\etat(\qbar)+h))\right]\right)\label{limit}\\
&=-\frac{\beta^2}{2}\left(1-\beta^2\E_{x^*}\left[\cosh^{-4}(\beta(\etat(\qbar)+h))\right]\right)\nonumber
\end{align}
where we have used the fact that $x^*_{m,r}\eval_{r=\qbar}=x^*$. Hence by the Taylor's expansion \eqref{taylor} and the equation \eqref{limit}, if $\beta$ and $h$ are such that
\begin{equation*}
\beta^2\E_{x^*}\left[\cosh^{-4}(\beta(\etat(\qbar)+h))\right]=\beta^2\int_\R \frac{e^{-z^2/2}}{\sqrt{2\pi}}\cosh^{-4}(\beta z\sqrt{\qbar}+h))>1,
\label{ATline}
\end{equation*}
we can pick $m$ close enough to $1$ and $r$ close enough to $\qbar$ so that the inequality \eqref{to prove AT} holds. 
This yields the desired sufficient condition for the high-temperature solution not to hold.
\appendix
 
\section{Topology on $\mc$}
\label{app:topo}
The weak, vague and weak-* convergences correspond on the space of probability measures $\mc$ as the measures are on a compact of $\R$. Moreover, any sequence is tight. Therefore, the weak topology is determined by the weak convergence. 

In studying the functionals on $\mc$, we are led to consider the topology on $\mc$ induced by the $L^1([0,1],g'(q)dq)$-norm on the distribution functions of the elements of $\mc$ where $g$ is a strictly increasing function in $C^1([0,1])$ with $g(0)=0$. It turns out that all these norms induce topologies on $\mc$ that are equivalent to the weak topology. 

First, we claim that the $L^1(g'(q)dq)$-topology is equivalent to the $L^1(dq)$-topology on the space $\mc$. 
Clearly, $\|\cdot\|_{L^1(g'(q)dq)}\leq\max_q g'(q)\|\cdot\|_{L^1(dq)}$. On the other hand, the following estimate holds for any $\delta>0$:
$$\|\cdot\|_{L^1(dq)}\leq \frac{1}{\delta}\|\cdot\|_{L^1(g'(q)dq)}+Leb\{q\in [0,1]:0\leq g'(q)<\delta\}$$
where $Leb$ stands for the Lebesgue measure. We use the fact that the distribution functions are bounded above by $1$ and below by $0$ to get the second term. As $g$ is strictly increasing, we have that, for any $\epsilon>0$, there exists $\delta(\epsilon)$ such that $Leb\{q\in [0,1]:0\leq g'(q)<\delta\}<\epsilon$. Let $x_\gamma$ be a net of distribution functions on $[0,1]$ that converges in the $L^1(g'(q)dq)$-norm. To see that $x_\gamma$ also converges in the $L^1(dq)$-norm, it suffices to see that for $\delta$ arbitrary small but fixed, one can also make the first term of the r.h.s. of the above estimate arbitrary small using the convergence in the $L^1(g'(q)dq)$-norm.

The equivalence with the weak topology is a direct consequence of the fact that the $L^1(dq)$-norm metrizes the weak topology on $\mc$ (see \cite{villani}).

\section{ Gaussian Differentiation Formulas}
\label{app:gaussian}
The differentiation of expectations of gaussian variables whose covariance depends on a parameter is facilitated by the following result which can be seen as an extension of the Wick's formula or simply gaussian integration by parts. 

\begin{prop}[\cite{ASS, ASS2}]
Consider a gaussian vector $\kappa=(\kappa_i, i\in\N)$ for which the covariance matrix $\{c_{ij}(t)\}$ depends on a parameter $t\in[0,1]
$. Assume $c_{ij}\in C^1([0,1]))$. We write $\E_t$ for the expectation over $\kappa$. Let $\phi:\R^\N\to\R$ in $C^2(\R^N)$ whose derivatives multiplied by $e^{-\epsilon |x|^2}$ are bounded functions for any $\epsilon>0$. Then
\begin{equation*}
\frac{d}{dt}\E_t[\phi(\kappa)]=\frac{1}{2}\sum_{i,j}c'_{ij}(t) \E_t\left[\partial_{\kappa_i}\partial_{\kappa_j}\phi(\kappa)\right]. 
\label{wick}
\end{equation*}
\end{prop} 

The proof of the proposition is easy to carry for polynomials (this case is the usual Wick's formula). For the general case, we refer to \cite{ASS2} for a proof using the Fourier transform.

In the case $\phi(\kappa)=\log\sum_\alpha\sa e^{\psi(\kappa_\alpha)}$ for a set of weight $\{\sa\}$, the formula becomes
\begin{align}
\frac{d}{dt}\E_t[\phi(\kappa)]&=\frac{1}{2}\sum_{\alpha}c'_{\alpha,\alpha}(t)\E_t\left[\psi''(\kappa_\alpha)+\psi'^2(\kappa_\alpha)\right]\frac{\sa e^{\psi(\kappa_\alpha)}}{\sum_{\alpha}\sa e^{\psi(\kappa_\alpha)} }
%\frac{1}{2}\frac{\sum_{\alpha}\sa e^{\psi(\kappa_\alpha)}c'_{\alpha,\alpha}(t)\E_t\left[\psi''(\kappa_\alpha)+\psi'^2(\kappa_\alpha)\right]}{\sum_{\alpha}\sa e^{\psi(\kappa_\alpha)} }
\label{wick 1st deriv}\\
&-\frac{1}{2}\sum_{\alpha,\alpha'}c'_{\alpha,\alpha'}(t)\E_t\left[\psi'(\kappa_\alpha)\psi'(\kappa_{\alpha'})\right]\frac{\sa\saprime e^{\psi(\kappa_\alpha)}e^{\psi(\kappa_{\alpha'})}}{\sum_{\alpha,\alpha'}\sa\saprime e^{\psi(\kappa_\alpha)}e^{\psi(\kappa_{\alpha'})}} .  \nonumber
%-\frac{1}{2}\sum_{\alpha,\alpha'}c'_{\alpha,\alpha'}(t)\E_t\left[\frac{\sum_{\alpha,\alpha'}\sa\saprime \psi'(\kappa_\alpha)\psi'(\kappa_{\alpha'})e^{\psi(\kappa_\alpha)}e^{\psi(\kappa_{\alpha'})}}{\sum_{\alpha,\alpha'}\sa\saprime}\right] .  \nonumber
\end{align}

It is possible to get higher-order derivatives by just applying the above gaussian differentiation formula successively. As an example, if the covariance $c_{ij}$ depends linearly on two parameters $s$ and $t$, $c_{ij}=c_{ij}(s,t)$, then applying the formula \eqref{wick} twice yields
\begin{equation}
\partial_s\partial_t \E_{s,t}[\phi(\kappa)]= \frac{1}{4}\sum_{i,j,k,l} \partial_s c_{ij}\partial_t c_{ij} \E_{s,t}\left[\partial_{\kappa_i}\partial_{\kappa_j}\partial_{\kappa_k}\partial_{\kappa_l}\phi(\kappa)\right].
\label{wick 2nd deriv}
\end{equation}


\begin{thebibliography}{10}

\bibitem{ALR} Aizenman M., Lebowitz J., Ruelle D., {\it Some rigorous results on the Sherrington-Kirkpatrick spin glass model}, Comm. Math. Phys. {\bf 112} (1987) pp. 3-20; 

\bibitem{ASS} Aizenman M., Sims R., Starr S., {\it An Extended Variational Principle for the SK Spin-Glass Model}, Phys. Rev. B {\bf 68} (2003), pp. 214403;

\bibitem{ASS2}Aizenman M., Sims R., Starr S., {\it Mean Field Spin Glass Models from the Cavity-ROSt Perspective}, preprint 2006, arXiv:math-ph/0607060;

\bibitem{BS}  Bolthausen E., Sznitman A.-S., {\it On Ruelle's Probability Cascades and an Abstract Cavity Method}, Comm. Math. Phys. {\bf 197} (1998) pp. 247-276; 

\bibitem{derrida} Derrida B., {\it Random-energy model: Limit of a family of disordered models}, Phys. Rev. Lett., {\bf 45} (1981) pp. 79-82; Derrida B, {\it Random-energy model: An exactly solvable model of disordered systems}, Phys. Rev. B, {\bf 24} (1981) pp. 2613-2626; Derrida B.  {\it A generalization of the random energy model which includes correlations between energies}, J. Phys. Lett. {\bf 46} (1985), pp. L401-L407;

\bibitem{Guerra_bound} Guerra F., {\it Broken Replica Symmetry Bounds in the Mean Field Spin Glass Model}, Comm. Math. Phys. {\bf 233} (2003) pp.1-12;

\bibitem{Guerra_cavity} Guerra F., {\it About the cavity fields in mean field spin glass models}, arXiv.org:cond-mat/0307673

\bibitem{Guerra_AT} Guerra F., {\it The replica symmetric region in the Sherrington-Kirkpatrick mean field spin glass model. The Almeida-Thouless line}, arXiv.org: cond-mat/0604674;

\bibitem{GT_quadratic}  Guerra F., Toninelli F. L., {\it Quadratic replica coupling in the Sherrington-Kirkpatrick mean field spin glass model}, J. Math. Phys. {\bf 43} (2002) pp. 3704;

\bibitem{Panchenko} Panchenko D., {\it A question about Parisi functional}, arXiv.org:/math.PR/0412463;

\bibitem{Ruelle} Ruelle D., {\it A Mathematical Reformulation of Derrida's REM and GREM},  Comm. Math. Phys. {\bf 108} (1987) pp. 225-239;

\bibitem{RA} Ruzmaikina A., Aizenman M., {\it Characterization of invariant measures at the leading edge for competing particle systems}, Ann. Probab. {\bf 33} (2005), pp.82Ð113;

\bibitem{zeproof} Talagrand M., {\it The Parisi Formula}, Ann. Math. {\bf 163} (2006), pp. 221-263;

\bibitem{Talagrand_parisi} Talagrand M., {\it Parisi measures}, J. Func. Anal., to appear;

\bibitem{AT} Toninelli F., {\it About the Almeida-Thouless transition line in the Sherrington-Kirkpatrick mean field spin glass model}, arXiv.org: cond-mat/0207296;

\bibitem{villani} Villani C., {\it Topics in Optimal Transportation}, AMS, Providence (2003), 370 pp.



% ., {\it },  {\bf } () p.;



\end{thebibliography}
\end{document}